\documentstyle[aps,epsfig,multicol]{revtex}

\begin{document}
\draft
\title{Fluctuations and correlations in hexagonal optical patterns}
\author{Dami\`a Gomila and Pere Colet}
\address{Institut Mediterrani d'Estudis Avan\c{c}ats, IMEDEA (CSIC-UIB),\\
Campus Universitat Illes Balears, E-07071 Palma de Mallorca, Spain.}
\maketitle

\begin{abstract}
We analyze the influence of noise in transverse hexagonal patterns in
nonlinear Kerr cavities. The near field fluctuations are determined by the
neutrally stable Goldstone modes associated to translational invariance
and by the weakly damped soft modes. However these modes do not contribute
to the far field intensity fluctuations which are dominated by damped
perturbations with the same wave vectors than the pattern. We find strong
correlations between the intensity fluctuations of any arbitrary pair of
wave vectors of the pattern.
Correlation between pairs forming $120^\circ$ is larger than between pairs
forming $180^\circ$, contrary to what a naive interpretation of emission
in terms of twin photons would suggest.
\end{abstract}

\begin{multicols} {2} 

\section{Introduction}

The properties of the fluctuations and correlations in spatially extended 
patterns out of thermodynamical equilibrium were studied long ago in the 
context of hydrodynamic systems \cite{Foster,Vinals}. More recently there 
has been a new surge of interest in the field of nonlinear and quantum optics 
since Lugiato and Castelli pointed out the existence of a purely quantum
phenomenon in a spatial stationary dissipative structure \cite{Lugiato}, the 
reduction of fluctuations below quantum limits in the difference between the 
intensities of the two Fourier modes of a stripe pattern. Since then, the 
properties of the fluctuations and correlations of stripe 
patterns in different nonlinear optical models have been widely studied
\cite{stripes,Miguelclass,Miguelquant,Roberta}. 
Stripe patterns in these systems appear as supercritical transitions, and 
close to the threshold for pattern formation the harmonics of the fundamental 
wave vectors can be neglected. Therefore considering the homogeneous mode and 
the two modes of the stripes for each field is enough and simplifies the 
problem allowing for an analytic treatment. 
The problem considering the whole infinite set of transverse modes in optical 
stripe patterns has been addressed numerically in \cite{Roberta}.
At difference with the few modes approximation, the continuous problem makes
evident the role of the so called soft modes in the fluctuations of the near
field. However a very good agreement between the few modes approximation and 
the continuous treatment is still found for the correlations of the 
fluctuations of the far field mode intensities close to threshold.

Due to the additional complexity, fluctuations and correlations in hexagonal 
optical patterns have been much less studied \cite{Grynberg,Alessandra}. 
Furthermore, 
these two papers use an approximation in few modes to describe the pattern. 
However hexagonal patterns generally appear sub-critically, with a finite
amplitude and therefore harmonics are not negligible even at threshold.
Strictly speaking, an approximation in which only the homogeneous mode plus 
the six modes of the hexagons are considered is not fully justified. 

In this paper we will treat the case of a subcritical hexagonal pattern using 
a continuous model, i.e., avoiding any restriction to a reduced number of 
spatial modes. In particular, we consider an optical cavity filled with a 
nonlinear isotropic Kerr medium. This situation is described by the 
Lugiato-Lefever model \cite{Lugiato-Lefever} in which a Turing instability 
was described for the self-focusing case. This pattern-forming instability 
leads to a hexagonal pattern in the 2D transverse plane. We find that the 
fluctuation of the near field are dominated by the two Goldstone modes 
associated to the translational invariance in the $x$ and $y$ directions, and 
by the soft modes arbitrarily close to them in infinitely large systems. 
We also show that the modes that dominate the near-field fluctuations do not 
contribute at all to the far-field intensity fluctuations. We identify the 
modes responsible for the far-field intensity correlations and we find: 
(a) strong correlations between arbitrary pairs of wave vectors of the 
pattern, but stronger between those forming a $120^\circ$ angle; 
(b) anti-correlation between the zero wave vector of the spectrum of 
fluctuations and any wave vector of the pattern. While the anti-correlation 
of the homogeneous mode with the off-axis wave vectors can be understood in 
terms of energy conservations, the common microscopic interpretation of the 
far-field intensity correlations in terms of emission of twin photons would 
naively suggest that the strongest correlation is between the wave vectors 
forming a $180^\circ$ angle. In fact, the total transverse momentum 
conservation involve always at least four modes simultaneously and give some 
hints about how the correlations should be, but does not identifies the 
pairs with stronger correlations. Our results here are obtained within a 
semi-classical approach in which specific features of quantum statistics are 
neglected.  

The paper is organized as follows: In Sec. \ref{model} we describe the model
we are considering. In Sec. \ref{method} we linearize around 
the hexagonal pattern and describe the linear response of the
system to noise perturbations. In Sec. \ref{near field} we discuss in detail
the field fluctuations, and in Sec. \ref{farfield} we
describe the correlations of the field Fourier components.
Finally, in Sec. \ref{conclusions} we give some concluding remarks.

\section{Description of the model}
\label{model}

The dynamics of the electric field inside an optical 
cavity with a self-focusing Kerr medium can be described, in the mean-field 
approximation, by a equation for the scaled slowly varying amplitude of the 
field $E(\vec{x})$ \cite{Lugiato-Lefever,Aguado} 
\begin{eqnarray}
\label{kerresc}
\partial_t E =-(1+i\theta)E+i\nabla^2 E+E_0 \nonumber \\
+i2 \vert E \vert^2 E+\xi (\vec{x},t) ,
\end{eqnarray}
where $E_0$ is the input field, $\theta$ is the cavity 
detuning, $\nabla^2$ is the transverse Laplacian and $\xi (\vec{x},t)$ is a 
complex Gaussian white noise with zero mean and correlations 
\begin{eqnarray}
\label{noise}
\langle \xi (\vec{x},t) \xi^* (\vec{x'},t')\rangle & = & \epsilon 
\delta(\vec{x}-\vec{x'})\delta(t-t') \nonumber \\
\langle \xi (\vec{x},t) \xi(\vec{x'},t')\rangle &= & 0.
\end{eqnarray}

In the absence of noise Eq.\ (\ref{kerresc})  has a homogeneous stationary 
solution $E_s$ given by
\begin{equation}
\label{homosol}
E_0=E_s[1 - i  (2 I_s - \theta)],
\end{equation}
where $I_s=\vert E_s \vert ^2$. It is well known that the homogeneous 
solution (\ref{homosol}) shows bistability for $\theta > \sqrt 3$. We will 
restrict ourselves to the non-bistable regime $\theta < \sqrt 3$. 

\begin{figure}	
\centerline{\epsfig{figure=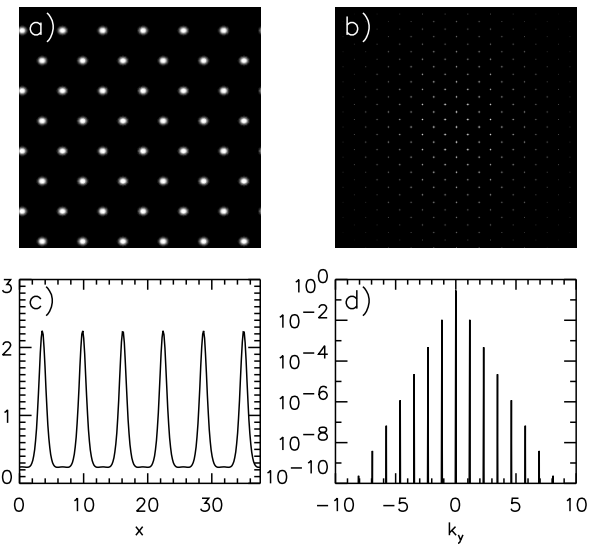}} 
\caption[]{(a) Near field intensity $|E(x,y)|^2$ and (b) power spectrum 
(far-field) $|E(\vec{k})|^2$ of a stationary hexagonal solution. Note the 
presence in the far-field of high order harmonics. (c) Shows a cross section 
along a $x$ axis of the intensity pattern and (d) a cross section along the 
$k_y$ axis of the power spectrum.}
\label{stationarysol}
\end{figure}

A linear stability analysis of the homogeneous solution with respect to
spatially periodic perturbations yields to the dispersion relation
\begin{equation}
\label{dispersion relation}
\lambda (\vec{k})=-1\pm \sqrt{-(\theta+k^2-6I_s)(\theta+k^2-2I_s)},
\end{equation}
where $\lambda(\vec{k})$ is the linear growth rate of a perturbation with 
wave vector $\vec{k}$ and $k=|\vec{k}|$. The instability threshold is located 
at $I_s^c = 1/2$ and the critical wave number is $k_c = \sqrt{-\theta + 2}$.
For pump intensities above threshold the maximum linear growth rate is for 
wave vectors with modulus
\begin{equation}
\label{kunstable}
k_u = \sqrt{-\theta + 4 I_s}.
\end{equation}
At threshold $k_u=k_c$. Starting from the homogeneous solution and changing 
the pump intensity to a value above, but close to, threshold a hexagonal 
pattern with a wave number $k$ close to $k_u$ arises (which as follows from 
Eqs. (\ref{kunstable}) and (\ref{homosol}) depends on the pump intensity). 
The transition is subcritical and the hexagonal pattern, once it is formed, 
is stable for values of the pump intensity within a quite large range which
includes values below threshold. Typically, the hexagons appear 
oscillating even at threshold.  The amplitude of the oscillations decreases 
decreasing the pump intensity until they become, in all cases, stationary. 
We are interested here in the properties of the fluctuations and correlations 
of the stationary hexagonal patterns. The oscillatory behavior of the 
hexagons is investigated in Ref. \cite{turbulence}.
Due to its subcriticality, the hexagonal pattern has always a finite 
amplitude, even at threshold. The harmonics of the six fundamental 
wave vectors have also a significant amplitude so that they have to be 
included in the calculations to obtained realistic quantitative results. 
This is particularly relevant here due to the self-focusing effect which 
leads to a pattern with high peaks and a strongly anharmonic far-field 
(See fig. \ref{stationarysol}). 

\section{Linearization around the hexagonal pattern}
\label{method}

\begin{figure}
\centerline {\psfig{figure=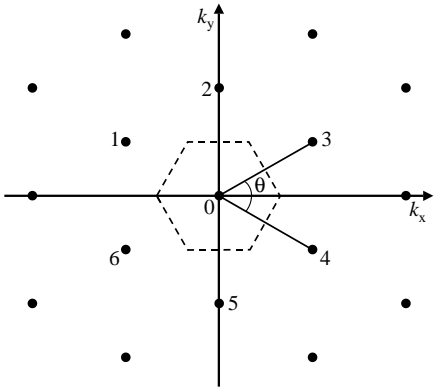,width=6cm,clip=true}}
\vspace{0.5cm}
\caption[]{First Brillouin zone (dashed hexagon) of the hexagonal lattice 
defined by the wave vectors of the pattern in the Fourier space.}
\label{brillouin}
\end{figure}

The stationary hexagonal pattern can be written in the form 
\begin{equation}
\label{solucio}
E_h(\vec{x})=\sum_{n=0}^N a_n e^{i \vec{k}^0_n (\vec{x}-\vec{x}_0)},
\end{equation}
where $a_n$ are complex coefficients, $\vec{x}_0$ determines the global 
position of the pattern (we take $\vec{x}_0=0$ in the following), 
$\vec{k}^0 _0$ is the homogeneous mode and $\vec{k}^0 _n$ for $n=1, \dots, N$ 
are the off-axis wave vectors of the  hexagonal pattern. Here we take $N=90$ 
which corresponds to considering up to the fifth order harmonics in the far 
field. The six fundamental harmonics have modulus $k_u$. For simplicity we 
first consider the case without noise.
Linearizing Eq. (\ref{kerresc}) around the stationary 
solution (\ref{solucio}) we obtain the following equation for the 
fluctuations $\delta E(\vec{x},t) = E(\vec{x},t) - E_h (\vec{x})$
\begin{eqnarray}
\label{eqfluct}
\partial_t \delta E &=& -(1 + i \theta ) \delta E + i \nabla^2 \delta E
\nonumber \\
&&+ i 2 [2 \vert E_h \vert ^2 \delta E +  E_h E_h \delta E^*].
\end{eqnarray}
As Eq. (\ref{eqfluct}) is a linear differential equation with periodic 
coefficients, a general bounded solution can be found under a Floquet form 
\cite{coullet}:
\begin{equation}
\label{floquetform}
\delta E(\vec{x},t) = \int e^{i \vec{q}\vec{x}} A(\vec{q},\vec{x},t) d\vec{q} 
\end{equation}
where $A(\vec{q},\vec{x},t)$ are functions with the same spatial
periodicity than the stationary pattern $E_h$, and therefore can be written as
\begin{equation}
A(\vec{q},\vec{x},t)=\sum_{n=0}^N \delta a_n(\vec{q},t)
e^{i\vec{k}^0_n\vec{x}}. 
\label{floquetform2}
\end{equation}
From (\ref{eqfluct}), (\ref{floquetform}) and (\ref{floquetform2}) we obtain
for each perturbation wave vector $\vec{q}$ a set of linear differential 
equations for the time evolution of the Fourier coefficients
$\delta a_n(\vec{q},t)$.
\begin{eqnarray}
\partial_t \delta a_n(\vec{q},t) =[-(1 + i \theta ) - i |\vec{k}^0_n 
+ \vec{q}|^2] \delta a_n(\vec{q},t) \nonumber \\
+ i2\{2\sum_{l=0}^N \sum_{m=0}^N a_l a_m^*\delta a_{n-l+m}(\vec{q},t)+ 
\nonumber \\
\sum_{l=0}^N \sum_{m=0}^N a_l a_m [\delta a_{-n+l+m}(-\vec{q},t)]^*\}  
\label{eqflucq}
\end{eqnarray}
where $\delta a_{n-l+m}(\vec{q},t)=\delta a_j(\vec{q},t)$ with 
$\vec{k}^0_j=\vec{k}^0_n-\vec{k}^0_l+\vec{k}^0_m$. Introducing 
$\vec{\Sigma}(\vec{q},t) = (Re[\delta a_0(\vec{q},t)]$, 
$Im[\delta a_0(\vec{q},t)]$, $\dots$, $Im[\delta a_N(\vec{q},t)]$,
$Re[\delta a_0(-\vec{q},t)]$, $\dots$, $Im[\delta a_N(-\vec{q},t)])^{\rm{T}}$,
Eqs.\ (\ref{eqflucq}) can be written as 
\begin{equation}
\label{eqflucq2}
\partial_t \vec{\Sigma}(\vec{q},t)= 
M(\vec{E}_h,\vec{q})\vec{\Sigma}(\vec{q},t).
\end{equation}
$\vec{\Sigma}(\vec{q},t)$ includes perturbations with $+\vec{q}$ and 
$-\vec{q}$ since they are coupled in Eq. (\ref{eqflucq}). The important point 
is that  perturbations with different $\vec{q}$ vectors are uncoupled. 
Any perturbation with a vector $\vec{q'}$ outside the first Brillouin zone
of the hexagonal lattice defined by the wave vectors of the pattern 
$\vec{k}^0 _n$ (See fig.\ \ref{brillouin}) is equivalent to another one with 
a vector $\vec{q}= \vec{q'} + \vec{k}^0_n$ inside.
 
Therefore the values of $\vec{q}$ to be considered are only those inside half 
of the first Brillouin zone.
In this way one finds a set of $4(N+1)$ eigenvalues and eigenvectors for 
each vector $\vec{q}$ except for the cases $\vec{q}=\vec{0}$ and 
$\vec{q}=\vec{k}_n^0/2$ where $\vec{\Sigma}(\vec{q},t)$ has only 
$2(N+1)$ components.  The eigenvalues may be either real or complex
conjugates and determine the linear response of the system when 
perturbations with $\vec{k}^0_n \pm \vec{q}$ wave vectors are applied. 

\begin{figure}
\centerline {\psfig{figure=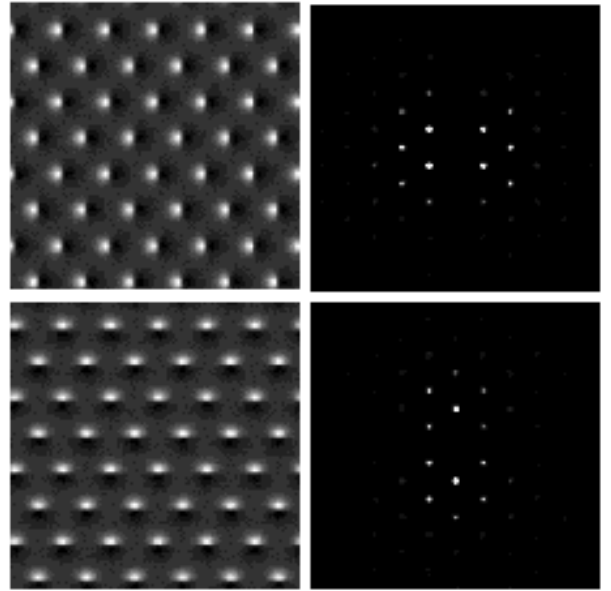,width=8cm,clip=true}}
\caption[]{Real part (left) and power spectrum (right) of the Goldstone modes 
$\partial_x E_h$ (top) and $\partial_y E_h$.}
\label{Goldstonemodes}
\end{figure}

\begin{figure}
\centerline {\psfig{figure=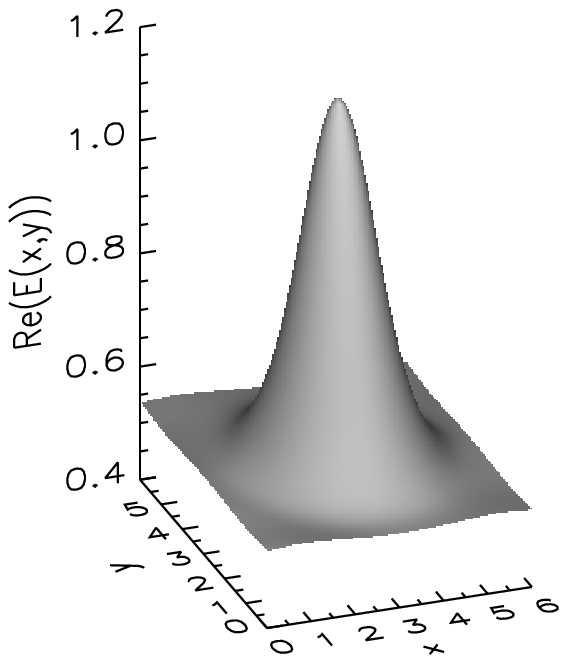,width=4cm}
\psfig{figure=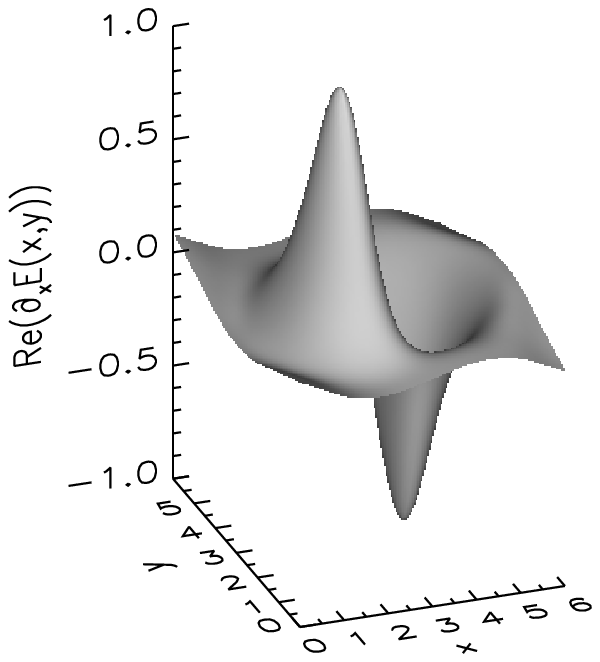,width=4cm}}
\caption[]{Close up of the real part of a peak of the hexagonal pattern 
(left) and the real part of the Goldstone mode $\partial_x E_h$ (right).}
\label{Goldstonemodes2}
\end{figure}

As the system is translationally invariant, $E_h(\vec{x}+\vec{x_0})$ is also 
a stationary solution for any fixed $\vec{x}_0$, the two modes 
$\partial_x E_h$ and $\partial_y E_h$ are eigenvectors of 
$M(\vec{E}_h,\vec{q}=0)$ with zero eigenvalue. These neutrally stable modes 
of the linearized dynamics are the so-called Goldstone modes 
\cite{Foster,Roberta} and correspond to homogeneous perturbations that 
displace rigidly the pattern along the $x$ or $y$ directions
\begin{eqnarray}
\partial_x E_h(\vec{x})&=&i\sum_n a_n k^0_{nx} e^{i\vec{k}^0_n\vec{x}} 
\label{gradient}\\
E_h(x,y)+x_0\partial_x E_h&=&
\sum_n a_n(1+ik^0_{nx}x_0) e^{i\vec{k}^0_n\vec{x}} 
\nonumber \\ \simeq \sum_n a_n e^{i\vec{k}_n(x+x_0,y)}&=&E_h(x+x_0,y).
\label{phasefluc}
\end{eqnarray}
For the hexagonal pattern considered here the Goldstone modes have the 
profile shown in Figs.\ \ref{Goldstonemodes} and \ref{Goldstonemodes2}. In 
the near field they have a larger amplitude at the borders of the hexagonal 
peaks, where the gradient is larger and show a sharp transition from negative 
to positive values just at the center of the peak (Fig. 
\ref{Goldstonemodes2}).
In a situation where the hexagonal pattern is stable the Goldstone modes are 
the only neutral modes while all the other eigenvectors have eigenvalues with 
strictly negative real part.

At this point we can study the linear response to noise perturbations.
Decomposing the noise term of (\ref{kerresc}) as it has been done for the
field [Eqs.\ (\ref{floquetform}) and (\ref{floquetform2})] we obtain an 
extra term in the rhs of equation (\ref{eqflucq2})
\begin{equation}
\label{eqflucq2noise}
\partial_t \vec{\Sigma}(\vec{q},t)= M(\vec{E}_h,\vec{q}) 
\vec{\Sigma}(\vec{q},t) +\vec{\Xi}(\vec{q},t),
\end{equation}
where $\vec{\Xi}(\vec{q},t)=(Re[\xi(\vec{k}_0^0+\vec{q}, t)]$, 
$Im[\xi(\vec{k}_0^0+\vec{q}, t)]$, $\dots$,
$Im[\xi(\vec{k}_N^0+\vec{q}, t)]$, $Re[\xi(\vec{k}_0^0-\vec{q}, t)]$, $\dots$,
$Im[\xi(\vec{k}_N^0-\vec{q}, t)])^{\rm{T}}$.
Then, we can write an Ornstein-Uhlenbeck process \cite{Gardiner} 
for the amplitude $\Theta_i(\vec{q},t)$ of each eigenvector of 
$M(\vec{E}_h,\vec{q})$
\begin{equation}
\label{O-U}
\partial_t\Theta_i(\vec{q},t)=\lambda_i(\vec{q})\Theta_i(\vec{q},t)
+\eta_i(\vec{q},t),
\end{equation}
where $\lambda_i(\vec{q})$ is the $i$-th eigenvalue of $M(\vec{E}_h,\vec{q})$
(ordered according to the value of its real part, 
$Re[\lambda_i(\vec{q})] \geq Re[\lambda_{i+1}(\vec{q})]$).
$\eta_i(\vec{q},t)= \sum_{n=0}^{4N+3} C^{-1}_{ij}(\vec{q})\Xi_j(\vec{q},t)$ 
is the noise expressed in the eigenvectors basis and $C(\vec{q})$ is the 
matrix for the change of basis as obtained diagonalizing $M(\vec{E}_h,
\vec{q})$. The coefficients $\delta a_n(\vec{q},t)$ are related to the 
amplitudes of the eigenmodes $\Theta_i(\vec{q},t)$ by
\begin{eqnarray}
Re[\delta a_n(\vec{q},t)]&=& \Sigma_i \Theta_i(\vec{q},t)C_{2n-1 i}
(\vec{q}) \nonumber  \\
Im[\delta a_n(\vec{q},t)]&=& \Sigma_i \Theta_i(\vec{q},t)C_{2n i}(\vec{q}) 
\nonumber \\
Re[\delta a_n(-\vec{q},t)]&=& \Sigma_i \Theta_i(\vec{q},t)C_{2n-1+2N i}
(\vec{q}) \nonumber  \\
Im[\delta a_n(-\vec{q},t)]&=& \Sigma_i \Theta_i(\vec{q},t)C_{2n+2N i} 
(\vec{q}).
\label{change of basis}
\end{eqnarray}  

The noises $\eta_i(\vec{q},t)$ are Gaussian and white in time and have 
cross-correlations
\begin{eqnarray}
\label{noisecorrelations}
\langle \eta_i(\vec{q},t) \eta_j^*(\vec{q'},t')\rangle &=&
{\epsilon \over 2}D_{i,j}(\vec{q})\delta(t-t')\delta(\vec{q}-\vec{q'}) \\
\langle \eta_i(\vec{q},t) \eta_j(\vec{q'},t')\rangle &=&
{\epsilon \over 2}\tilde{D}_{i,j}
(\vec{q})\delta(t-t')\delta(\vec{q}-\vec{q'}).
\end{eqnarray}
where 
\begin{eqnarray}
\label{Dmatrix}
D_{i,j}(\vec{q})=\sum_{k=0}^{4N+3} C^{-1}_{ik}(\vec{q})C^{-1*}_{jk}(\vec{q})
\nonumber \\
\tilde{D}_{i,j}(\vec{q})=\sum_{k=0}^{4N+3}
C^{-1}_{ik}(\vec{q})C^{-1}_{jk}(\vec{q}).
\end{eqnarray}

The solution of the stochastic process is then
\begin{equation}
\Theta_i(\vec{q},t)=e^{\lambda_i(\vec{q})t}\int_0^t e^{-\lambda_i(\vec{q})s}
\eta_i(\vec{q},s) ds.
\label{O-U solution}
\end{equation}
The average value of the eigenmodes amplitude is $\langle 
\Theta_i(\vec{q},t)\rangle = 0$.
The correlations between the amplitudes of the eigenvectors are given by:
\begin{eqnarray}
\langle \Theta_i(\vec{q},t)\Theta_j^*(\vec{q'},t') \rangle = 
{ \epsilon D_{ij}(\vec{q})\over 
-4(\lambda_i(\vec{q})+ \lambda_j^*(\vec{q'}))}\nonumber \\
(1 - e^{(\lambda_i(\vec{q})+\lambda_j^*(\vec{q'}))t}) \delta(\vec{q}-\vec{q'})
\label{O-U solution1} \\
\langle \Theta_i(\vec{q},t)\Theta_j(\vec{q'},t') \rangle = 
{ \epsilon  \tilde{D}_{ij}(\vec{q})\over 
-4(\lambda_i(\vec{q})+ \lambda_j(\vec{q'}))}\nonumber \\(1 - 
e^{(\lambda_i(\vec{q})+\lambda_j(\vec{q'}))t})\delta(\vec{q}-\vec{q'})
\label{O-U solution2}
\end{eqnarray}
In particular, the time evolution of the mean value of the squared amplitude 
of the eigenvectors with non zero eigenvalue is
\begin{equation}
\langle |\Theta_i(\vec{q},t)|^2 \rangle  = { \epsilon 
D_{ii}(\vec{q}) \over -8Re[\lambda_i(\vec{q})]} (1 - 
e^{2Re[\lambda_i(\vec{q})]t}).
\label{meanvalue}
\end{equation}
For times much longer than a characteristic time $\tau_i(\vec{q})\sim 
-1/Re[\lambda_i(\vec{q})]$ this mean squared amplitude reaches a 
stationary value 
\begin{equation}
\label{stationarystate}
\langle |\Theta_i(\vec{q})|^2 \rangle = { \epsilon D_{ii}(\vec{q})\over 
-8Re[\lambda_i(\vec{q})]}.
\end{equation} 
Eq.\ (\ref{stationarystate}) does not apply to the Goldstone modes as they 
have zero eigenvalue ($\lambda_0(\vec{q}=\vec{0})=0$). Its time evolution is 
given by (from Eq.\ (\ref{O-U solution}))
\begin{equation}
\Theta_i(\vec{q},t)=\int_0^t \eta_i(\vec{q},s) ds.
\label{O-U Goldstone}
\end{equation}
This is a purely diffusive motion which never reaches a stationary 
state. Its mean squared amplitude grows linearly in time
\begin{equation}
\label{Goldstonediff}
\langle |\Theta_0(\vec{q}=\vec{0},t)|^2 \rangle ={\epsilon
\over 8}D_{00}(\vec{q}=0)t.
\end{equation}
Therefore, the linearization fails for times 
$t \sim 1/\epsilon$, when the amplitude of the Goldstone modes 
$|\Theta_0(\vec{q}=\vec{0})|^2$ reaches values comparable to one and nonlinear
terms will come into play.

\section{Fluctuations in the near field}
\label{near field}

\begin{figure}	
\centerline{\psfig{figure=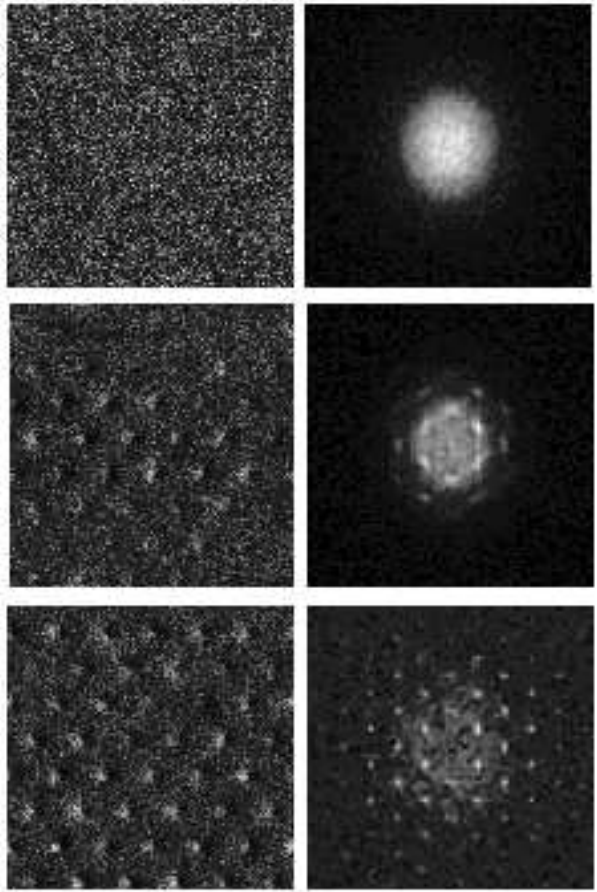,width=8cm,clip=true}} 
\caption[]{Near field (left: $Re[\delta E(\vec{x})]$, 
$\delta E(\vec{x})=E(\vec{x})-\langle E(\vec{x})\rangle$) and far field 
(right: $|\delta E(\vec{k})|$, 
$\delta E(\vec{k})=E(\vec{k})-\langle E(\vec{k})\rangle$) pattern 
fluctuations after switching on the noise. From top to bottom: $t=2, t=200$, 
and $t=2000$. We have considered $\epsilon=10^{-6}$.}
\label{fluctuations}
\end{figure}

Starting from a stationary hexagonal solution of the system equations without
noise, a typical evolution of the pattern fluctuations when the noise is
switched on, obtained from numerical integration of the nonlinear equations 
Eq.\ (\ref{kerresc}) \cite{numerics}, is shown in  Fig. \ref{fluctuations}.
 
From Eq. (\ref{meanvalue}) we get that for short time all the modes 
$\Theta_i(\vec{q},t)$ have a similar mean squared amplitude which is
proportional to the noise intensity and grows linearly with time  
\begin{equation}
\langle |\Theta_i(\vec{q},t)|^2 \rangle  = {\epsilon \over 4}
D_{ii}(\vec{q}) t .
\label{shortime}
\end{equation}
All the modes will contribute with a similar weight to the field
fluctuations and therefore there is a complete lack of structure in the 
field fluctuations at short times as shown in Fig. \ref{fluctuations} (top).

\begin{figure}
\centerline{\epsfig{figure=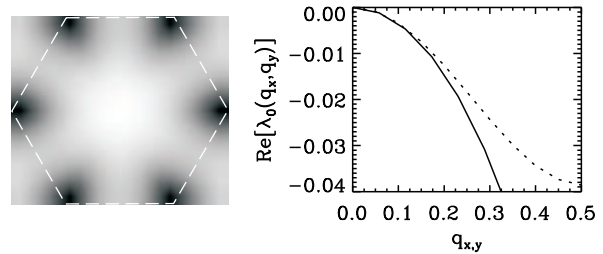}}
\caption[]{Left: $Re[\lambda_0(\vec{q})]$,  
darker color indicates smaller values. The white line shows the 
first Brillouin zone. The center of the figure corresponds
to the Goldstone modes ($Re[\lambda_0(\vec{q}=0)]=0$). Right: transverse cut
of  $Re[\lambda_0(\vec{q})]$ along $q_x$ (solid line) and $q_y$ (dotted
line) axis.}
\label{lambdaq}
\end{figure}

\begin{figure}
\centerline {\psfig{figure=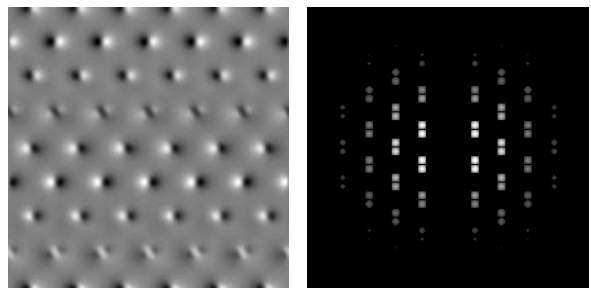,width=8cm,clip=true}}
\caption[]{Soft mode $\vec{q}=(0, 2\pi/L_y)$ whose decay rate 
is $Re[\lambda(\vec{q}=(0,2\pi/L_y))]=-0.0052$.}
\label{soft mode}
\end{figure}

As time goes on, following (\ref{meanvalue}), the mean squared amplitude of 
the modes does not grow linearly any more. It reaches a steady state value
given by (\ref{stationarystate}) which is larger for the modes which have a
smaller decaying rate $Re[\lambda_i(\vec{q})]$. 
According to the way the eigenvalues has been ordered, for a given $\vec{q}$, 
the smaller decaying rate is $Re[\lambda_0(\vec{q})]$. The eigenmodes with
eigenvalue $\lambda_0(\vec{q})$ are the so-called soft modes 
\cite{Foster,Roberta,Ma} which are connected with the Goldstone modes and 
for which $\lambda_0(\vec{q} \gtrsim \vec{0}) \sim -|\vec{q}|^2$ (see Fig.
\ref{lambdaq}).
While the Goldstone modes correspond to neutrally stable homogeneous 
perturbations, the soft modes correspond to weakly damped long-wavelength 
perturbations. For systems with a finite size $L$ the less 
damped of the soft modes are the ones with the smallest $\vec{q}$ vector
allowed by the size of the system, $|\vec{q}|=2\pi/L$. These modes have a 
decay rate proportional to $1/L^2$ and therefore a stationary mean squared 
amplitude proportional to $\epsilon L^2$.
The shape of one soft mode is illustrated in Fig. \ref{soft mode}.
Its profile is similar to a Goldstone mode but with a long-wavelength 
modulation on top of it.
The wavevector of the modulation is precisely the wavevector $\vec{q}$ that 
identifies the soft mode.
While the Goldstone modes describe an overall rigid motion of the pattern,
the soft modes add some distortion so the pattern moves in slightly different
form at different locations. 

For the numerical simulations shown in Fig.\ \ref{fluctuations}, at $t=200$ 
the mean squared amplitude of most of the modes has already saturated at 
small values, while the amplitude of the soft modes is, in average, just 
reaching the stationary value.
Therefore, as shown in the figure (center) the field fluctuations are 
dominated by the soft and Goldstone modes and short range spatial structures 
start to appear. The typical correlation length of the fluctuations is 
determined by the wavelength of the soft modes.

According to (\ref{Goldstonediff}) the mean square amplitude of the Goldstone 
mode keeps growing linearly in time and does not saturate in the linear
approximation. The Goldstone mode amplitude only saturates due to 
nonlinearities at times of the order $t \sim 1/\epsilon$ where the linear 
theory for fluctuations described in the previous section fails (this time
is around $10^6$ for the simulation shown in Fig. \ref{fluctuations})
Assuming linearization is valid, at times $t \gg L^2$, the
mean squared amplitudes of the Goldstone modes ($\sim \epsilon t$, 
(\ref{Goldstonediff})) become much larger than the amplitude of any of the 
soft modes and dominate the fluctuations. In our system at time $t \sim 2000$ 
the profile of the field fluctuations is already determined only by the 
Goldstone modes as shown in Fig. \ref{fluctuations} (bottom). In the far 
field the largest fluctuations are those of the wave vectors of the pattern, 
while in the near field fluctuations show a long range structure with strong 
correlation over all the system size. Integrating for longer times will not 
change the profile of the field fluctuations, the amplitude of the Goldstone
mode will eventually saturate but nevertheless will completely determine the 
field fluctuations.

For systems with a larger system size, the soft modes will take a longer time
to reach the stationary amplitude (and in fact its amplitude will be larger).
However, provided that $L^2<1/\epsilon$, the evolution will be basically the
same, showing the three stages discussed before.
If the system size is larger than $1/\sqrt{\epsilon}$ then the 
stationary amplitude of the soft modes and the Goldstone modes will be
determined by the system nonlinearities and if $L$ is large enough, both 
amplitudes may have similar sizes, in which case Goldstone and soft modes will
contribute to  the near field fluctuations even at $t \rightarrow \infty$ 
and therefore the pattern of fluctuations may show a finite correlation length
in the stationary regime.

\section{Fluctuations and correlations in the far field}
\label{farfield}

In the following three subsections, we address the fluctuations in Fourier 
space of the field, intensity and momentum.

\subsection{Field fluctuations}

In the Fourier space the field fluctuations are also dominated by the 
Goldstone and soft modes. From Eqs.\ (\ref{gradient}) and (\ref{phasefluc}), 
we can see how the Goldstone modes, which have the same wave vectors than the 
hexagonal pattern, induce opposite and very large phase fluctuations in 
opposite Fourier components, which correspond to the rigid translation of the 
hexagons in the near field. The homogeneous component of the field is not 
affected by the Goldstone modes. The soft modes do not have exactly the same 
Fourier components than the hexagonal pattern
but they are very close, therefore their main contribution is to
broaden the spots of the far-field fluctuations (see the far-field for the
intermediate time in Fig. \ref{fluctuations}).
 
\subsection{Intensity fluctuations and correlations}

The intensity fluctuations of the far-field peaks are 
\begin{equation}
\delta I(\vec{k})=I(\vec{k})-I_h(\vec{k}),
\label{Ifluc}
\end{equation}
where $I(\vec{k})=|E(\vec{k})|^2$ and $I_h(\vec{k})=|E_h(\vec{k})|^2$.
The correlation function of the far-field intensity fluctuations of a 
fundamental wave vector of the pattern, for instance $\vec{k}_3^0$, with the 
far-field intensity fluctuations of any other wave vector $\vec{k}$ is given 
by 
\begin{equation}
C_1(\vec{k}_3^0,\vec{k})={\langle \delta I(\vec{k}_3^0) 
\delta I(\vec{k})\rangle
\over \sqrt{\langle |\delta I(\vec{k}_3^0)|^2 \rangle \langle |\delta 
I(\vec{k})|^2 \rangle}}.
\label{corrI} 
\end{equation}

\begin{figure}
\centerline {\psfig{figure=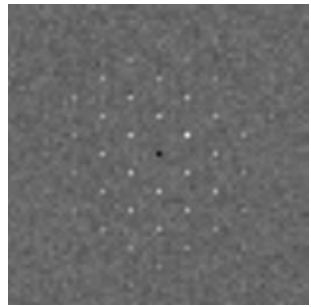,width=4cm}}
\vspace{0.5cm}
\caption[]{Correlation function of the intensity fluctuations 
$C_1(\vec{k}^0_3, \vec{k})$. The brightest spot takes the value $1$ 
corresponding to the autocorrelation of $\vec{k}^0_3$. Note the strong 
correlation with the other six fundamental wave vectors of the pattern and 
with the first rings of harmonics. The correlation decay as one consider 
higher and higher harmonics. Note also the anti-correlation with the 
homogeneous as a black spot at the center. The grey background is around $0$}
\label{intensitycorr}
\end{figure}

\begin{figure}
\centerline {\psfig{figure=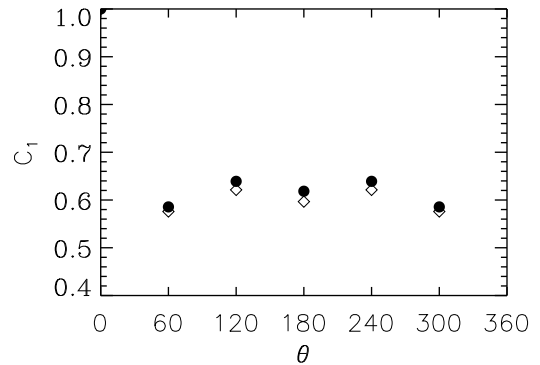,width=8cm}}
\caption[]{Correlations between the intensity fluctuations of the fundamental
wave vectors obtained from the  numerical integration of equation 
(\ref{kerresc})(rhombi), and analytically form Eq.\ (\ref{analyticalC1}) 
(dots).}
\label{c1analytic}
\end{figure}

From the numerical integration of Eq.\ (\ref{kerresc}),and  averaging over 
two hundred realizations of the noise \cite{numerics},
we find strong correlations between the intensity fluctuations of all the 
modes of the pattern, not only among the fundamental harmonics but also with 
the higher order ones (Fig. \ref{intensitycorr}). 
For the fundamental harmonics, 
the correlation $C_1(\vec{k}_3^0,\vec{k}_i^0)=\langle \delta I(\vec{k}_3^0)
\delta I(\vec{k}_i^0)\rangle$, with $i=1,6$, is larger for the modes forming 
an angle $\theta={2\pi \over 3}$ and smaller for $\theta={\pi \over 3}$ 
(Fig. \ref{c1analytic}). Using a 6 modes approximation
\cite{Grynberg,Alessandra} one finds that $N_i+N_{i+1}-N_{i+3}-N_{i+4}$ 
(being $N_i$ the number of photons of the mode $i$ ) is a conserved quantity 
for the interaction Hamiltonian related to momentum conservation. This means 
that one should expect strong correlations among these sets of four Fourier 
modes. However this reasoning can not reveal which modes within these sets 
are more correlated in pairs. So, the stronger correlation between those 
modes forming a $\theta={2\pi \over 3}$ angle, despite fulfill momentum 
conservation, can not be completely understood in this terms. 
We also find strong anti-correlations between the intensity fluctuations of 
the modes of the pattern and the homogeneous mode which are related to energy 
conservation (Fig. \ref{intensitycorr}). 

\begin{figure}
\centerline {\psfig{figure=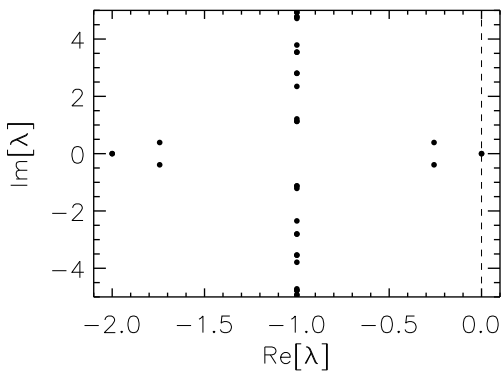,width=8cm}}
\caption[]{Eigenvalues of the eigenvectors with $\vec{q}=\vec{0}$. There is 
two zero eigenvalues corresponding to the Goldstone modes. Note also the 
symmetry of the spectrum with respect to the axis $Re[\lambda]=-1$.}
\label{someigvalues}
\end{figure}

Neglecting terms of order $\epsilon^2$, the far-field intensity 
fluctuations can be approximated as
\begin{equation}
\delta I(\vec{k})\approx 2Re[E_h^*(\vec{k})\delta E(\vec{k})],
\label{Ifluc1storder}
\end{equation}
where $\delta E(\vec{k})=E(\vec{k})-E_h(\vec{k})$. As $E_h(\vec{k})=
\sum_n a_n (2\pi)^2 \delta(\vec{k}^0_n -\vec{k})$ we have to consider only 
perturbations that have the same wave vectors of 
the pattern ($\vec{q}=\vec{0}$). Therefore, the soft modes ($\vec{q}\gtrsim 
\vec{0}$) do not contribute to the fluctuations of the far-field
intensity peaks. 
The Goldstone modes are indeed $\vec{q}=0$ perturbations, but
the intensity fluctuations associated to the Goldstone modes are 
$\delta I(\vec{k})=2 Re[E_s(\vec{k})^*\partial_x E_s(\vec{k})]$ which from
(\ref{solucio}) and (\ref{gradient}) is exactly zero. So, neither the 
Goldstone modes do contribute to the fluctuations of the far-field intensity 
peaks. They only contribute to phase fluctuations. Therefore, fluctuations of 
the far-field intensity peaks have to be described by the other eigenvectors 
of $M(E_h,\vec{q}=\vec{0})$. These eigenvectors have the same Fourier 
components than the hexagonal pattern and their eigenvalues 
$\lambda_i(\vec{q}=\vec{0})$, 
($i=1,\dots N$) are shown in Fig. \ref{someigvalues}.

Using Eqs.\ (\ref{change of basis}), (\ref{O-U solution1}) and 
(\ref{O-U solution2}) we can compute the correlation function 
$C_1(\vec{k}^0_n,\vec{k}^0_m)$ analytically
\begin{eqnarray}
C_1(\vec{k}^0_n,\vec{k}^0_m)=  2 Re[&E_h^*&(\vec{k}^0_n)	
E_h^*(\vec{k}^0_m) \langle \delta E(\vec{k}^0_n) \delta E(\vec{k}^0_m)
\rangle + \nonumber \\
&E_h&(\vec{k}^0_n) E_h^*(\vec{k}^0_m)\langle \delta E(\vec{k}^0_n) 
\delta E^*(\vec{k}^0_m)\rangle,
\label{analyticalC1}
\end{eqnarray}
where
\begin{eqnarray}
\label{correlation2}
&\langle \delta E(\vec{k}^0_n)\delta E(\vec{k}^0_m)\rangle = &\nonumber \\ 
&\sum_i \sum_j {\epsilon \over -4(\lambda_i + 
\lambda_j)}(1-e^{(\lambda_i +\lambda_j)t})\tilde{D}_{ij}(\vec{q}=0)&
\nonumber \\ &(C_{2n-1 i}C_{2m-1 j}-C_{2n i}C_{2m j}+ &
\nonumber \\ &i(C_{2m-1 i}C_{2n j}+C_{2n-1 i} C_{2m k}))& \\
&\langle \delta E(\vec{k}^0_n) \delta E^*(\vec{k}^0_m)\rangle =& \nonumber \\ 
&\sum_i \sum_j {\epsilon \over -4(\lambda_i + 
\lambda_j^*)}(1-e^{(\lambda_i + \lambda_j^*)t}) D_{ij}(\vec{q}=0)&
\nonumber \\ &(C_{2n-1 i}C_{2m-1 j}+C_{2n i}C_{2m j}+ &
\nonumber \\ &i(C_{2m-1 i}C_{2n j}-C_{2n-1 i} C_{2m k}))&
\label{corrC3}
\end{eqnarray}
Fig. \ref{c1analytic} shows the stationary value ($t \rightarrow \infty$) of 
(\ref{analyticalC1}) for the fundamental wave vectors of the pattern
as a function of the angle $\theta$ between them. The results obtained from
the linearized theory are in very good agreement with the correlations 
obtained from the numerical simulations of the nonlinear equation 
(\ref{kerresc}).

\begin{figure}
\centerline {\psfig{figure=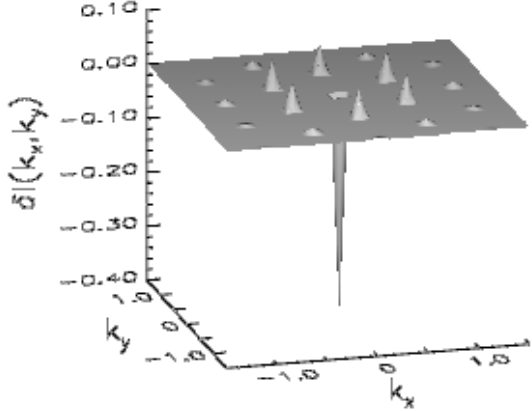,width=8cm}}
\caption[]{Intensity fluctuations due to the eigenmodes with 
$Re[\lambda]=-0.25$}
\label{modecorrenergy}
\end{figure}

\begin{figure}
\centerline {\psfig{figure=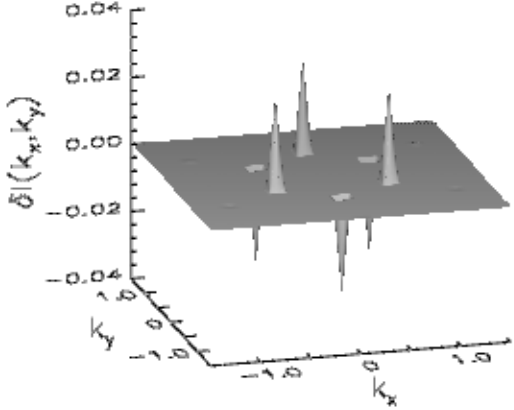,width=8cm}}
\caption[]{Intensity fluctuations due to one of the eigenmodes with 
$Re[\lambda]=-1$}
\label{modecorrmomentum}
\end{figure}

From Fig. \ref{someigvalues} we can see that the most important 
eigenvectors are those associated to the complex conjugate eigenvalues 
with $Re[\lambda_2(\vec{q}=0)]= Re[\lambda_3(\vec{q}=0)]= -0.25$. This pair 
of eigenvectors give the strong correlation between all the Fourier 
components and a strong anti-correlation with the homogeneous field. The 
excitation of this eigenvectors implies also anti-correlations between the 
homogeneous mode and the off-axis Fourier components of the pattern (Fig. 
\ref{modecorrenergy}).
Some eigenvectors with $Re[\lambda_n(\vec{q}=0)]=-1$ are finally the 
responsible for the differences between the correlations in the fluctuations 
of the six firth harmonics. The typical profile of one of these eigenmodes is 
shown in Fig. \ref{modecorrmomentum}. The symmetry of the three peaks
located at $120^\circ$ yields to larger correlations between the fluctuations 
of the corresponding far field intensity peaks. 

We note that the modes that determine the correlations of the far-field 
intensity peaks reach stationary values in a much shorter 
time ($t\sim -1/Re[\lambda_2(\vec{q}=0)]=4$)
than the Goldstone and soft modes contributing to the near field 
fluctuations ($t\sim 1/\epsilon \sim 10^6$ as discussed in the previous 
section). As these correlations are determined by $\vec{q}=0$ perturbations,
these calculations can be performed in systems with relatively small size.
Increasing the system size, may change the near field profile of the 
fluctuations, as discussed before, but nevertheless the dynamical evolution 
of the modes contributing to the far-field intensity fluctuations and their 
mean squared stationary amplitude will be basically the same.

\subsection{Transverse momentum fluctuations}

\begin{figure}
\centerline{\psfig{figure=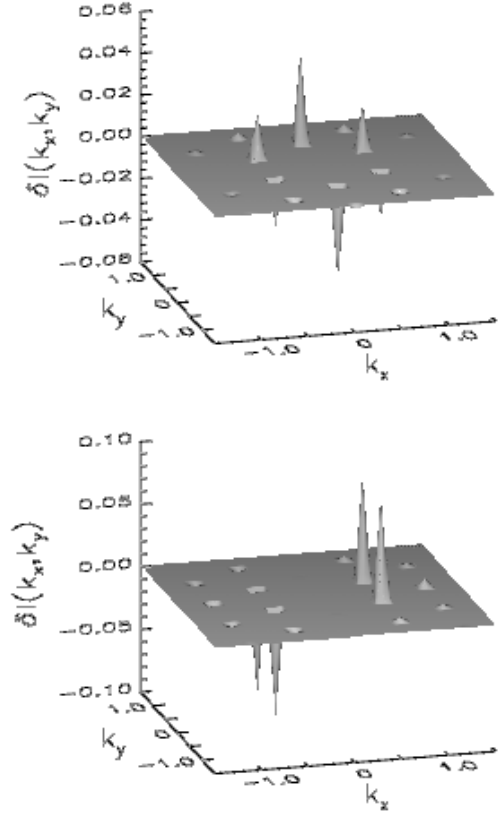,width=8cm}}
\caption[]{Intensity fluctuations due to the two modes
$Re[\lambda_{2N}(\vec{q}=0)]=Re[\lambda_{2N-1}(\vec{q}=0)]=-2$.}
\label{modemoment}
\end{figure}

We finally address the fluctuations of the transverse momentum. Without noise 
the total transverse momentum of the pattern is 
$\vec{P}=\sum_n |a_n|^2 \vec{k}_n^0=0$.
The noise induce fluctuations given by $\delta\vec{P}=\sum_n 2 Re[a_n^* 
\delta a_n] \vec{k}_n^0$.
One finds that all the eigenvectors that contribute to the far-field intensity
fluctuations strictly fulfill momentum conservation except for 
two modes with $\lambda_{2N}(q=0)=\lambda_{2N-1}(q=0)=-2$, which are exactly 
those symmetric to the Goldstone modes respect to the line $Re[\lambda]=-1$
(Fig. \ref{someigvalues}).
Therefore momentum fluctuations are determined by the two modes with 
maximum damping which are shown in Fig.\ \ref{modemoment}, the on the top
figure breaks the conservation of the $P_y$ momentum component while the 
one in the bottom breaks the conservation of $P_x$.
Alternatively it is possible to see from the classical field theory that 
the momentum fluctuations are damped with a coefficient -2. 
A Lagrange density can be defined for Eq.\ (\ref{kerresc}) \cite{lagrange}
\begin{eqnarray}
\label{lagrangeden}
L=e^{2t} [a|\nabla E|^2 + {i \over 2}(E\dot{E^*} -\dot{E}E^*)+ 
\nonumber \\
i(E_0E^*-E_0^*E)+ \theta |E|^2- |E|^4 ]
\end{eqnarray}
Substituting (\ref{solucio}) in (\ref{lagrangeden}) and integrating over all 
the space we get a Lagrangian for the amplitude of the modes
$a_n$ and for the position of the rolls in the near field $\vec{x_0}$
\begin{eqnarray}
\label{lagrangian}
&L=e^{2t} [\sum_n |a_n|^2 |\vec{k_n}|^2 +{i \over 2}\sum_n 
(\dot{a_n^*}a_n -& \nonumber \\
&\dot{a}_n a_n^*-i2|a_n|^2 
\vec{k_n}\dot{\vec{x}_0})+ i(E_0a_0^*-E_0^*a_0)+& \nonumber \\
&\sum_n |a_n|^2 - \sum_n \sum_{n'} \sum_m  a_n a_{n'} a_m^* a_{n+n'+m}^*].&
\end{eqnarray}
$\vec{x}_0$ is a cyclic coordinate, so its conjugate momentum 
\begin{equation}
\label{conjmomentum}
{\partial L \over \partial \dot{\vec{x}_0}}= 
-i2e^{2t}\sum_n |a_n|^2\vec{k_n} = -2e^{2t}\vec{P}
\end{equation}
is constant. Therefore $\dot{\vec{P}}=-2 \vec{P}$. $\vec{P}$ is identically 
zero without noise. When noise is present momentum fluctuations should 
satisfy 
\begin{equation}
\label{momentumfluctuations}
\dot{\delta \vec{P}}= -2 \delta\vec{P} + \chi(t),
\end{equation}
where $\chi(t)$ is a Gaussian white noise. Therefore, momentum fluctuations 
have the maximum damping. The same damping coefficient for the momentum 
fluctuations has been found by Gatti and Mancini from a few modes quantum 
formulation \cite{Alessandra}.

\section{Conclusions}
\label{conclusions}

We have analyzed the fluctuations and correlations in a hexagonal pattern of a
prototypical model in nonlinear optics. In the near field fluctuations are
dominated by the neutrally stable Goldstone modes associated to the 
translational invariance as well as by the soft modes connected with them.
The soft modes destroy the long range correlation in the fluctuations, however
in small systems these modes reach an stationary amplitude much earlier 
(and at a smaller value) than the Goldstone mode so that they are important 
only at intermediate times. At long times the fluctuations are dominated by 
the Goldstone modes which correspond to rigid displacements of the overall 
pattern. For very large systems, both the Goldstone and soft modes may have 
similar amplitudes and contribute to the fluctuations.

In the far-field, the most relevant effect of noise are the intensity 
fluctuations of the Fourier modes of the hexagonal pattern. At first order in
noise intensify, these fluctuations are not affected neither by the 
Goldstone modes nor by the soft modes. They are dominated by damped modes, 
so they reach stationary values in relatively short times. Their main 
characteristics are:
i) strong correlations between the intensity fluctuations of any arbitrary 
pair of the six fundamental wave 
vectors of the pattern, and also with their higher harmonics, ii) larger 
correlation between intensity fluctuations of the Fourier modes forming 
$120^\circ$ angles than between modes forming $180^\circ$, and iii) strong 
anti-correlations between the zero wave vector and the pattern Fourier modes.
Finally only the eigenmodes with maximum damping contribute to the
fluctuations of the total transverse momentum, therefore the total transverse 
momentum has the least possible fluctuations. 

Our results are obtained from both semianalytical calculations based on 
linearization around the hexagonal pattern and from numerical simulations of 
the nonlinear system. Some of our results and predictions are very general, 
and depend only on basic symmetry properties of the system such as 
translational invariance. Thus we expect that similar structure and 
properties of the fluctuations and correlations can be found in other 
nonlinear systems displaying hexagonal patterns.     

From a computational point of view, correlations in the far field intensity
peaks are practically independent of the system size and therefore can be 
calculated accurately in relatively small systems, provided all the relevant 
harmonics of the pattern are considered. Good statistics can be obtained 
integrating the nonlinear equations over relatively short times (even though 
the near field fluctuations are quite far away from reaching a stationary 
value). Of course, alternatively a linear semianalytical approach as the 
one described in Sec. \ref{method} can also be used to calculate far-field 
correlations.

The authors acknowledge helpful discussions with M. San Miguel and R. 
Zambrini, and financial support from the EC TMR Network QSTRUCT 
(FMRXCT960077) and from the MCyT (Spain, Projects PB97-0141-C02-02, 
BFM2000-1108 and BFM2001-0341-C02-02).

\end{multicols}
\end{document}